# XMM-Newton (cross)-calibration

M.G.F. Kirsch[*a], B. Altieri[a], B. Chen[a], F. Haberl[b], L. Metcalfe[a], A.M.T. Pollock[a], A.M. Read[c], R.D. Saxton[a], S. Sembay[c], M.J.S. Smith[a]

[a]European Space Agency (ESA), European Space Astronomy Center (ESAC),
Villafranca, Apartado 50727, 28080 Madrid, Spain
[b]Max-Planck-Institut für extraterrestrische Physik, Giessenbachstrasse 1, 85784 Garching, Germany
[c]Dept. of Physics and Astronomy, Leicester University, Leicester LE1 7RH, U.K.

## ABSTRACT

ESA's large X-ray space observatory XMM-Newton is in its fifth year of operations. We give a general overview of the status of calibration of the five X-ray instruments and the Optical Monitor. A main point of interest in the last year became the cross-calibration between the instruments. A cross-calibration campaign started at the XMM-Newton Science Operation Centre at the European Space Astronomy Centre in collaboration with the Instrument Principle Investigators provides a first systematic comparison of the X-ray instruments EPIC and RGS for various kind of sources making also an initial assessment in cross calibration with other X-ray observatories.

## 1. INTRODUCTION

XMM-Newton[1] was launched in December 1999 with an Ariane 504 rocket from French Guyana and operates six instruments in parallel on its 48-hour highly elliptical orbit. Three Wolter type 1 telescopes with 58 nested mirror shells focus X-ray photons on the five X-ray instruments of the European Photon Imaging[2,3] Camera (EPIC) and the Reflecting Grating Spectrometers[4] (RGS). In addition a 30 cm Ritchey Chrétien optical telescope is used for optical observations in parallel with the Optical Monitor[5] (OM). EPIC consists of two parts: EPIC-MOS (Metal-Oxide Semi-conductor) and EPIC-pn (p-n-junction). The two EPIC-MOS cameras use front illuminated MOS-CCDs as X-ray detectors while the EPIC-pn camera is equipped with a pn-CCD, which has been specially developed for XMM-Newton. EPIC provides spatially resolved spectroscopy over a field-of-view of 30' with moderate energy resolution. The EPIC camera can be operated in different observation modes related to the different readouts in each mode. For a detailed description of the modes see Kendziorra et al. (1997)[6], Kendziorra et al. (1999)[7], Kuster et al. (1999)[8] and Ehle et al. (2003)[9]. The RGS is designed for high-resolution spectroscopy of bright sources in the energy range from 0.3 to 2.1 keV. The OM extends the spectral coverage of XMM-Newton into the UV and optical, and thus opens the possibility to test models against data over a broad energy band. Six filters allow colour discrimination, and there are two grisms, one in the UV and one in the optical, to provide low resolution spectroscopy.

After reaching a good status for the calibration of the individual instruments (see sections 2 - 4) it is now time to focus strongly on the cross-calibration of the XMM-Newton instruments. Currently a special campaign is being carried out by the XMM-Newton Science Operations Centre in collaboration with the PI teams. The first step in this campaign has been to perform a systematic data analysis in order to identify and quantify cross-calibration deficiencies. Later suggestions shall be made in order to improve the cross calibration by refining the physical knowledge of all three instruments. This also includes a cross-calibration exercise with the Chandra observatory, however is mainly focussed for the time being on XMM-Newton internal agreement.

[*] mkirsch@xmm.vilspa.esa.es  +34 91 8131 345; fax +34 91 8131 172

## 2. STATUS OF EPIC CALIBRATION

In the following section we give a short summary of the current status of the EPIC calibration as implemented in SAS 6.0 with all available CCF at 30/06/2004, highlighting the major calibration efforts of the last year. Furthermore the outlook is considered for improvements of calibration that at the moment can be expected for the next SAS and/or CCF releases. The detailed status of calibration is given by Kirsch[10].

### 2.1 Astrometry

The precision with which astronomical coordinates can be assigned to source images in the EPIC focal plane is called Astrometry. We distinguish between the Absolute Astrometry (relative to optical coordinates, without taking into account possible shifts due to spacecraft miss-pointing), the Relative Astrometry per camera (within one camera after applying possible shifts due to spacecraft miss-pointing) and the Relative Astrometry between cameras (positions in one camera relative to an other one). The XMM-Newton absolute astrometry accuracy is limited by the precision of the Attitude Measurement System. Fig. 1 shows that the shift from the XMM-Newton to the optical frame is on average 0'' with a standard deviation of less than 0.8'' per axis. Hence the Absolute Pointing Accuracy is considered to be better than 1.0" (r.m.s.). The relative astrometry within each camera is accurate to 1.5" for all cameras and over the full field of view. The MOS metrology has been revised with SAS6.0, searching for systematics in the offsets of MOS peripherical CCDs with respect to the central CCD by using observations on rich stellar fields. CCD offsets of up to 2.7" have been corrected in the LINCCORD CCF issue 17. With this new CCF the MOS relative astrometry accuracy has been assessed to be 1.5" (r.m.s.) while it is as good as ~1.0" (r.m.s.) for EPIC-pn. Among all three EPIC cameras the relative astrometry is also estimated to be better than 1.5" across the whole field of view. Note that for faint MOS sources near the detection limit the statistical accuracy of the measurement limits the 90 % confidence contours to 2-4". A possible residual in the position angle rotation (Euler ? angle) of the order of 0.1 deg is under investigation. This could lead to an uncertainty of up to 1.5'' at the edge of the XMM-Newton field of view.

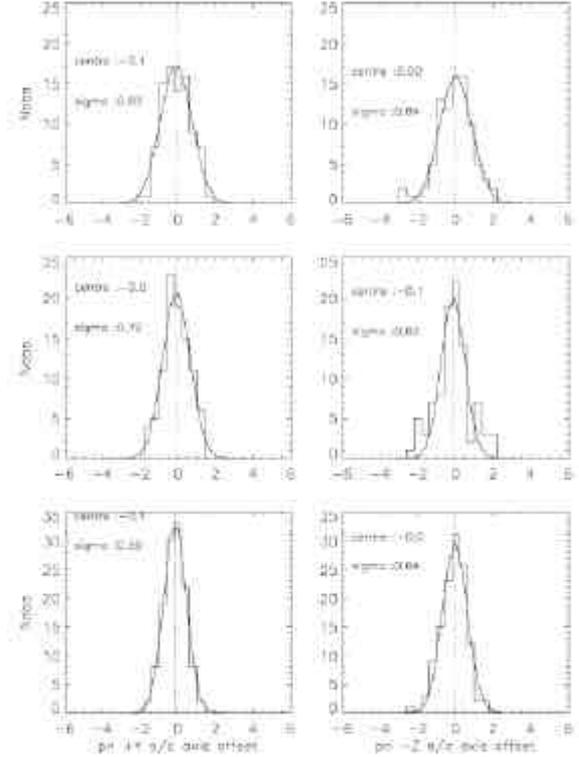

Figure 1:
Histogram of the distribution of offsets for each EPIC camera with respect to the 2MASS reference frame projected on to the two spacecraft axis for MOS1 (upper) MOS2 (middle) and pn (lower)
(For details see XMM-SOC-CAL-SRN-168)

### 2.2 Point Spread Function

The Point Spread Function is the spatial distribution of light in the focal plane in response to a (monochromatic) point source. The PSF integrates to 1 over the focal plane. Though the shape of the PSF is quite complex, the radially averaged profile can be adequately represented by an analytic function - a King function - whose parameters, core radius $r_0$ and index $a$, are themselves functions of energy and off-axis angle:

$$PSF = A \left[ 1 + \left( \frac{r}{r_0} \right)^2 \right]^{-a}$$

It is worth noting that both this function and its integral are analytical. Earlier work (EPIC-MCT-TN-011, EPIC-MCT-TN-012, XMM-CCF-REL-116) had used many bright point sources both on and off axis to determine the energy dependent PSF. This resulted in a linear dependence of $r_0$ and $a$ with energy and off-axis angle. It is shown in XMM-SOC-CAL-SRN-167 that this linear dependence is inaccurate - the dependencies of $r_0$ and $a$ are seen to be flatter (almost constant) with energy (at least out to ~ 8-10 keV). Thereafter the dependencies rapidly turn steeper.

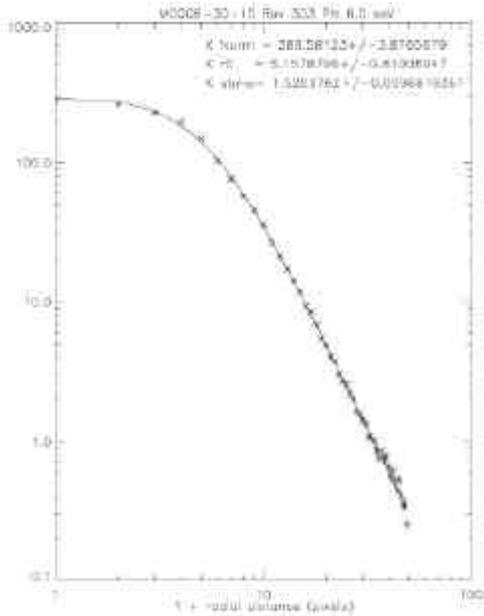

Figure 2: Surface brightness radial profile (crosses) plus fitted King profiles (lines) for MCG-6-30-15 Rev. 303 pn at 6 keV.

Two threads of analysis using data from very long and clean SW mode observations of very bright non piled-up sources were followed: One involved the forming of narrow-energy-band images, and fitting the surface brightness radial profiles obtained from these images with a King function to obtain $r_0$ and $a$ as a function of energy. A second analysis thread involved the extraction of spectra from narrow annuli around point sources, and once ARF files had been generated (this involving the actual form of the PSF), the spectra were fitted with standard spectral models, to see how (if at all) the spectral parameters obtained varied with extraction radius. This whole process was repeated for several sets of PSF parameters (including those obtained from the surface brightness radial profile fitting described above). An example of surface brightness radial profile plus fitted King profile is shown in Fig. 2.

The resultant dependencies of $r_0$ and $a$ are seen to be flatter (almost constant) with energy, at least up to ~ 8-10 keV (where the $r_0$ and $a$ relationships turn over) than in the previous parameterisation of the PSFs.

The new PSFs were used in the analysis of spectra extracted from narrow annuli around a number of bright point sources, as described above. A point source, of course, should show no variation in fitted spectral parameters whether the spectrum is extracted from the very centre of the distribution or from the wings, but use of previous PSFs result in a range in spectral parameters for different radii. Use of the new PSFs gives rise to very significant improvement, with the fitted normalization and power-law index remaining constant and independent of extraction radius.

A major problem with the previous parameterisation was its inability to produce consistent spectral fits for annular extraction regions such as those used for instance for the analysis of piled-up sources. Hence MCG-6-30-15 (from Rev 302) has been extracted from annuli of 5-40'', 10-50'' and 15-60'' and fits compared to those of a circular extraction (0-30''). This has been performed using the new and the old PSFs, and the results are presented in Fig. 3. Whereas usage of the old PSFs results in a per instrument normalization variation of up to 40%, and changes in the fitted spectral slope of 0.2, the new PSFs give rise to normalization variations of nearer 5% and a spectral slope change of at most 0.03.

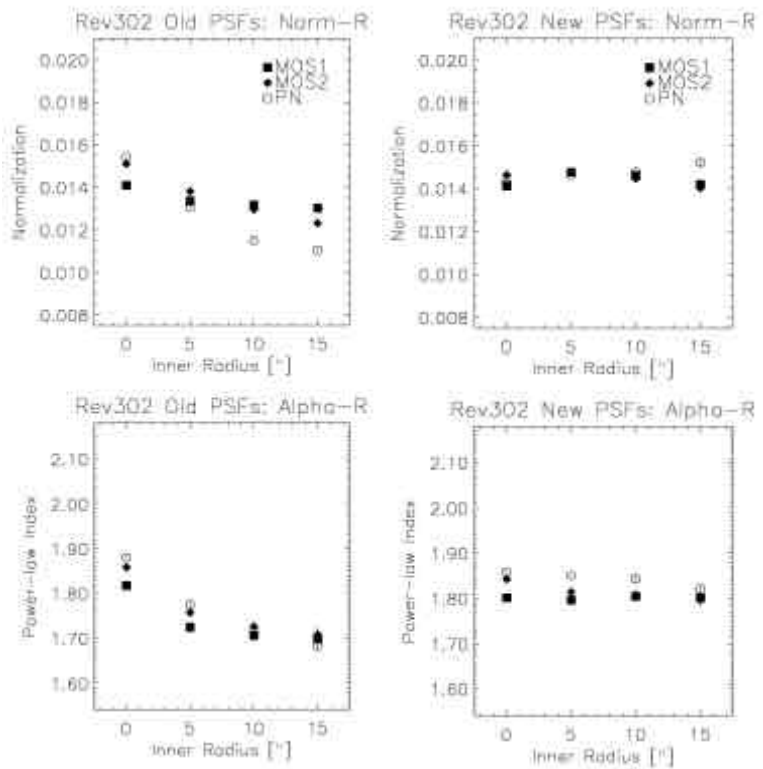

Figure 3: Plots showing how the fitted normalization (top) and power-law index (bottom) vary as a function of extraction region (left to right: 0-30" circle, 5-40" annulus, 10-50" annulus, 15-60" annulus), using the old CCF PSFs (left) and the new CCF PSFs (right) for the MCG-06-30-15 Rev.302 data.

Note that the King function is a good but not perfect fit to the PSF of the telescopes, as the core of the PSF is very slightly underestimated. This affects the MOS more than the pn (as the MOS detector pixels are much smaller than the pn pixels). This can produce an error in the enclosed energy of at most ~2 %, depending on instrument, energy and extraction radius. Work is currently underway to model the PSF as a combination of a King function plus a Gaussian function (the latter to model the slight excess at the core).

As yet, no sources bright enough for this new type of analysis to be performed off-axis have been observed. As such, the general off-axis results of previous work (EPIC-MCT-TN-011, EPIC-MCT-TN-012, XMM-CCF-REL-116) have been used to transform the new on-axis parameters presented here to projected off-axis values.

While the EPIC-pn PSF is azimuthally symmetric, the placing of the CCDs in the MOS cameras to follow the focal plane results in a chip-to-chip variation in the MOS PSF[11]. This is not currently modelled in the SAS calibration but will result in an azimuthal variation in the encircled energy fraction that is dependent on the extraction radius and the off-axis angle. This variation is not yet quantified but is estimated to be ± 4% for a source circle of radius 25'' at an off-axis angle of 7'.

**2.3 Vignetting**

Vignetting means the reduction in the effective area with radial distance from the telescope's axis. One of the most important outstanding problems of the calibration was an offset of around 1' in the telescope axis from nominal. This did not affect the astrometry but could have been the reason for some of the flux discrepancies between MOS and pn caused by the vignetting correction, which has not yet been adapted to this offset in SAS versions earlier than 6.0.0. The offset was determined and implemented in the corresponding CCF (XMM_MISCDATA_0020). With SAS 6.0.0 and the current available CCF the new consideration of the right optical axis position improves the vignetting correction. However the vignetting correction itself has not changed at all, the only difference is, that it is now applied for correct off axis angles, that could not be calculated correctly before due to the wrong information for the optical axis. This improves differences in flux for off axis sources for each camera from ± 14 % to ± 5 %.

**2.4 Energy Redistribution**

Energy Redistribution means**:** The energy profile recorded by the detector system in response to a monochromatic input.

**MOS Low Energy:** Observations of calibration targets confirm a significant change in the low energy redistribution characteristics of the MOS cameras with time. This change is probably due to an increase in the surface charge loss property of the CCDs that degrades the low energy resolution. Epoch dependant calibration files have been produced which reflect these changes, but users should be aware that uncertainties in the model of the redistribution function of the MOS cameras remain. Spectral fitting can be performed down to 150 eV, but in these cases it is recommended that a systematic error of 2 % be applied. A detailed description is given by S. Sembay et al.[12].

**PN redistribution:** EPIC-pn spectra from ? Puppis have shown that the spectral response below about 400 eV is not yet correctly reproduced. In particular the re-distribution as modelled in SAS 6.0.0 is higher than seen in the data. This can lead to large (30%) systematic errors in the absolute flux of very soft spectral components (kT<100 eV). Further observations with different read-out modes are planned to investigate the problem.

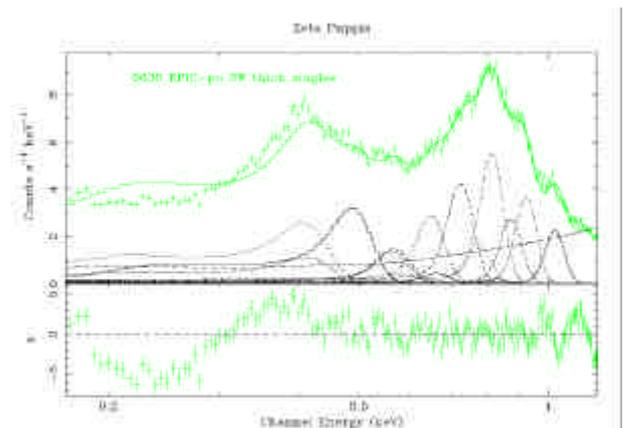

Figure 4: ? Puppis EPIC pn with redistribution SAS6.0

**2.5 Energy Calibration**

Charge Transfer Inefficiency is the imperfect transfer of charge as it is transported through the CCD to the output amplifiers during read-out whereas Gain is the conversion (amplification) of the charge signal deposited by a detected photon, from ADU (Analogue to digital unit) charge into energy (electron-volts). An improvement in the epoch dependent CTI and Gain correction in SAS 6.0.0 has reduced the uncertainty in the energy calibration from 10 to 5 eV for the imaging modes of the MOS cameras. (See: XMM-SOC-CAL-SRN-161). For MOS-Timing mode the CTI correction was changed improving over correction by debugging some erroneous code in SAS. MOS Timing mode energy accuracy does now agree with the imaging modes within 0.3 % (see Fig.5).

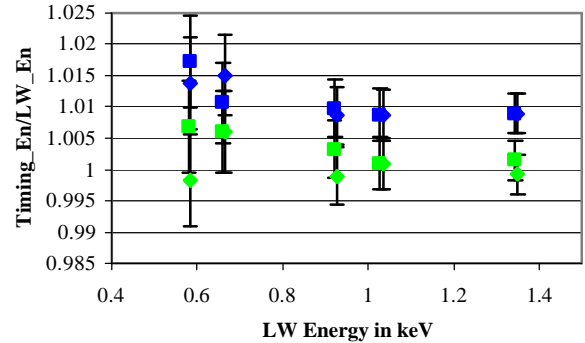

Figure 5: Line positions ratio of MOS LW and MOS Tining mode for of the SNR 1E0102-7219 before and after the correct CTI application.

The EPIC-pn Small Window mode showed a Gain/CTI under correction of ~2-3 % most prominent around the Oxygen-edge. This could lead to residuals in the fitted spectra of up to 20 %. A better CTI correction will be provided as soon as possible with a new CTI CCF issue. (See: XMM-SOC-CAL-SRN-162).

The internal calibration source shows an over correction of up to 15 eV at Mn-K in pn Extended Full Frame mode, that is related to imperfect CTI correction. This is currently under investigation with special calibration observations.

**2.6 New features in SAS 6.0**

The new SAS task *epreject* corrects shifts in the energy scale of specific pixels due to high-energy particles hitting the EPIC pn detector during offset map calculation and suppresses the detector noise at low energies by statistically flagging events based on the known noise properties of the lowest energy channels. In the case of timing mode data, flagging of soft flare events may be performed. An additional function is added to *emevents* that performs a blanking of bad-energy columns and handles those now correctly as dead areas. *emevents* is now doing a filtering & removal of flickering pixels decreasing significantly the noise at low-energies. The new SAS tasks *ebadpixupdate* allows operations on bad pixels at the level of the calibrated events list. The optional input of a background eventset to *epatplot* now allows the determination of background-subtracted pattern fractions. This is useful, e.g., in the case of extended source analysis or close to spectral background features. *badpixfind* now permits bad pixel searching on calibrated multi-chip (i.e. final pipeline product) eventsets. Previous versions only operated on raw event sets.

## 3. STATUS OF RGS CALIBRATION

**3.1 The Stability of the RGS**

About halfway through its four years in operation so far, in common with the MOS instruments described above, the operating temperature of the focal plane detectors of both Reflection Grating Spectrometers was decreased by about 30K in 2002 November giving a dramatic reduction in the number of noisy hot CCD pixels and columns. The behaviour of the CCDs in flight is described elsewhere in these proceedings[13].

Otherwise, the responses of both RGS1 and RGS2 have remained stable throughout the mission as described in detail in the latest calibration status document[14]. For example, Fig. 6 shows the relative flux observed in soft, medium and hard energy bands below the silicon edge of the SNR 1ES0102-7219 in RGS1 up until the end of the fourth year of operations. The measurements of this constant source are stable at levels that vary between 2 % and 4 % in the different bands, consistent with statistical uncertainties.

This stability is also evident in other aspects of the performance. For example, accumulation of blank exposures at a variety of levels of the local space environment's particle radiation allows reliable independent estimates of the instrumental background to be synthesized for the majority of observations. This is especially useful for extended sources that fill up much or all of the RGS field-of-view and thus preclude the standard off-axis point-source background methods. The SNR data in Fig. 6 are a case in point where the stability of the flux estimates is partly due to the quality of the background estimates which make up about 10 % of the total detected count rate. It is planned to make these methods publicly available by the end of 2004. The RGS stability is further demonstrated by the combination of several exposures of individual line-rich stellar coronae during routine calibrations that reveals the presence of many weak features in the spectra that are masked by statistical fluctuations in single observations. The rigorous combination of RGS spectra from different observations should be supported in a forthcoming release of the XMM data analysis tools

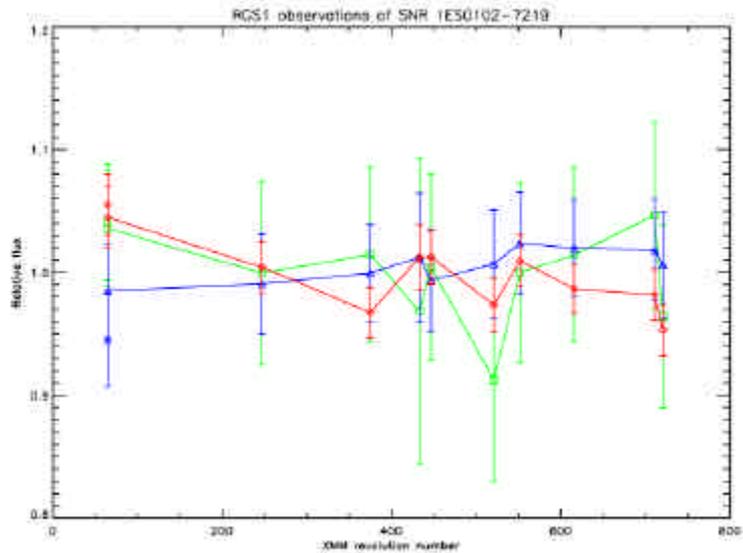

Figure 6: Relative flux observed in soft (red), medium (blue) and hard (green) energy bands below the silicon edge of the SNR 1ES0102-7219 in RGS1 up until the end of the fourth year of operations.

**3.2 The RGS Wavelength Scale**

The RGS wavelength scale derived by comparison of measured positions of narrow lines with their laboratory values[15] shows typical 1σ errors of about 7 mÅ. While this is a relatively small fraction of the instrumental line width it is clear that the errors are dominated by systematic components whose origin is under investigation. When brought under control, wavelength measurements should approach the level of the statistical errors which with the RGS sensitivity can be as small as 1 mÅ.

**3.3 Small-scale features in RGS spectra**

An important scientific problem facing some RGS observers is the identification of cosmic absorption features in the otherwise smooth continuum spectra of some AGN. Such features tend to be of low equivalent width and thus care is needed to describe accurately any complicating instrumental lines. The RGS detectors show well-documented Oxygen absorption near 23 Å in the effective area model and a weaker feature from Fluorine near 18 Å is taken into account in the latest calibration release. In work of this type, comparison of RGS1 and RGS2 is a worthwhile check, where possible, in order to identify low-level misbehaviour of mildly cool or warm CCD pixels or columns. Recent progress has brought a degree of control over systematic errors of this type with the introduction in SAS v6 of the choice of applying individual pixel offset corrections as an alternative to the usual node-based corrections.

## 4. STATUS OF OM CALIBRATION

In the following section we give a short summary of the current status of the OM calibration as implemented in SAS 6.0 with all available CCFS at 30/06/2004, highlighting the major calibration efforts of the last year. The detailed status of calibration is given by Chen[16].

## 4.1 Zero points and in-flight throughput

The zero points of the OM instrument are based on simulations of the Vega count rates, (i.e., Vega has zero magnitude in OM filters). The throughput is based on several spectrophotometric standards. Fig. 7 shows the count rates of OM in three UV filters as a function of time, indicating about 10 % degradation during XMM-Newton 4.5 year operations. For the three optical filters (UBV), the degradation is smaller, about 5 % over the whole time.

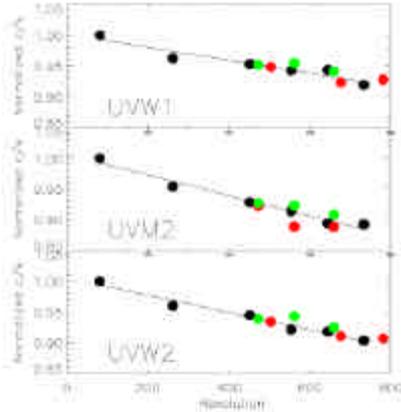

Figure 7: The normalized count rates of OM as a function of time in UVW1, UVM2 and UVW2 filters, for three white dwarfs, BPM16274 (black), Hz 2 (red), GD153 (green).

## 4.2 OM Photometry

**4.2.1 Colour transformation:** The UBV colour transformations from the OM instrumental system to Johnson's system were established based on observations. Several fields have been observed from the ground with the standard UBV filters and with the XMM-Newton OM. 363 cross-identified stars have been used to make the colour transformation (Antokhin et al. 2002). Currently, the colour-transformations for UV filters are based on the simulations because we have not got enough calibration observations for UV filters.

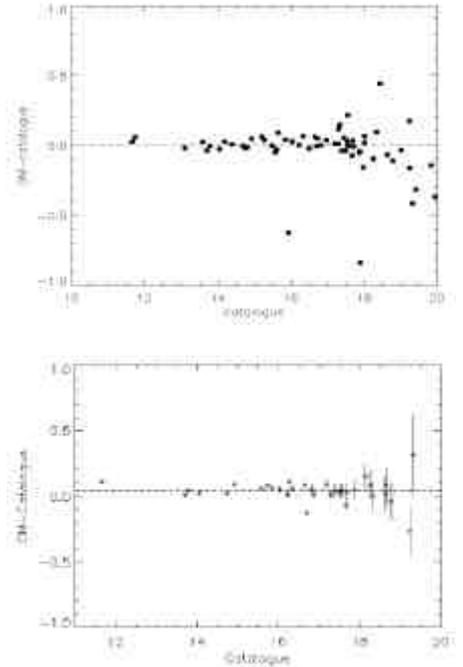

**4.2.2 Testing OM photometry with external catalogues:** A Landolt standard star field (SA95) has been observed with OM at rev. 407 and rev. 759. Fig. 8 shows the difference in standard V magnitude between the OM measurement and Stetson's measurement. From Fig. 8, we can see that the OM magnitude is in very good agreement with the Stetson catalogue magnitude in rev. 407, but there is an offset of 0.04 mag in rev. 759. This can be explained as the degradation as shown in Fig. 7. A time-dependent count rate correction will be added into the Current calibration files (CCF). Since these stars have very high photometric accuracy in the Stetson's catalogue, thus providing a direct measurement of the OM photometric accuracy. We found that the OM photometric accuracy is 2.3 %.

Figure 8: The difference in standard V magnitude between Stetson's measurement and OM measurement for Landolt standard star field SA95 in rev. 407 (upper) and rev. 759 (lower).

Fig. 9 shows UV colour transformation from two observations, SA95 field (asteroids), and G153 field (diamonds). The dashed line is the colour transformation in the CCF based on the simulations. We plan to observe a field with many blue stars with B-V < 0.4, then we will update the UV colour transformation based on observations.

## 4.3 OM count rate to flux conversion

We have derived the flux conversion factors from five white dwarfs (HZ4, GD50, HZ2, GD153, and G93-48). For each filter, if you multiply the count rates (counts/s) by the following numbers, you will get the flux (erg/cm$^2$/s/A).

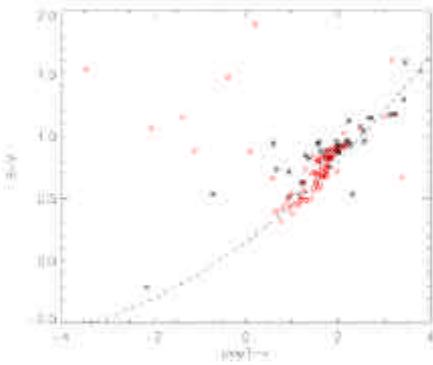

Figure 9: Comparing the observed UV color-transformation for two fields, SA95 (red) and G153 (black), with the simulated one (dashed line).

|                       | V        | B        | U        | UVW1     | UVM2     | UVW2     |
|-----------------------|----------|----------|----------|----------|----------|----------|
| Flux conversion Factor| 2.178E-16| 1.041E-16| 1.776E-16| 4.396E-16| 2.027E-15| 5.797E-15|
| Effective Wavelength (nm) | 543  | 434      | 344      | 291      | 231      | 212      |

We should point out, these flux conversion numbers provide an approximate measurement of the flux densities without *a priori* knowledge of the spectral type. Work on the spectral type dependent conversion factors is in progress.

### 4.4 Grisms

Several spectrophotometric stars and F-type stars have been observed to establish the wavelength and flux calibration of both Visible and UV grisms. This new version of SAS 6.0 contains for the first time tools to extract and calibrate the spectra produced by OM grisms automatically. The main source of error in the grism wavelength scale is due to the determination of the centroid of the zero order, which can produce a shift of 10 Angstroms. The accuracy of the grisms flux calibration is about 10 %.

## 5. CROSS-CALIBRATION

In 2003 significant manpower was added to the cross-calibration efforts of the PI instrument teams by the XMM-Newton Science Operations Centre (SOC). A detailed report will soon be available on the XMM-Newton calibration portal[17]. For various types of sources like AGN, SNR and stars XMM-observations have been selected out of the XMM-Newton Science Archive and were processed with the public SAS extracting source and background spectra and the respective response and effective area matrices. Spectral models have been built with XSPEC v11.3. For smooth continuum sources, the model was initially based on EPIC data and subsequently fitted simultaneously with the RGS. For line-dominated sources such as coronae and SNRs a phenomenological model has first been built of continuum and as many lines as needed to reproduce well the RGS before fitting simultaneously with EPIC. A further approach for some sources was to use the RGS fluxed spectrum as a model input.

The normalisation of all instruments is done with an additional multiplicative constant where the pn constant is set to 1. In XSPEC terminology: *model=const*(any_number_of_model_components)*

The sources in Table 1 have been used for the cross calibration. We will show representative results for some of the source classes.

All data have been used for sources whose count rates are constant within 5 % in an observation. For variable sources strict simultaneity was required. For both RGS and EPIC, all data falling on the sky projection of the RGS detectors were selected for extended sources. The following six energy bands have been adopted to investigate the energy dependence of the normalisation between instruments: 0.35-0.7 keV, 0.7-0.972 keV, 0.972-1.84 keV, 1.84-3.0 keV, 3.0-6.0 keV, 6.0-12.0 keV. The limit of 0.35 keV was chosen as the lowest energy from the RGS response, 0.972 keV is halfway between the strong lines of NeIX at 0.9220 keV (13.4471 Å) and NeX at 1.0220 keV (12.1321 Å).

Table 1: cross-calibration Targets

| AGN          | Stars       | SNR         | Neutron Stars   |
|--------------|-------------|-------------|-----------------|
| PKS2155-304  | AB Dor      | 1ES0101-7219| RX J0720.4-3125 |
| 3C273        | Capella     | N132D       | RX J0806.4-4123 |
| PKS0558-504  | YY Gem      |             |                 |
| Mkn 509      | UX Ari      |             |                 |
| Mkn 180      | zeta Puppis |             |                 |
| PG 0050+124  |             |             |                 |
| PG 1116+215  |             |             |                 |
| MCG-6-30-15  |             |             |                 |

## 5.1 AGN

The normalisation factors with respect to pn in the energy bands defined above are shown in Fig. 10 for different AGNs, as function of energy: MOS1 in red, MOS2 in green, RGS1 in dark blue and RGS2 in light blue

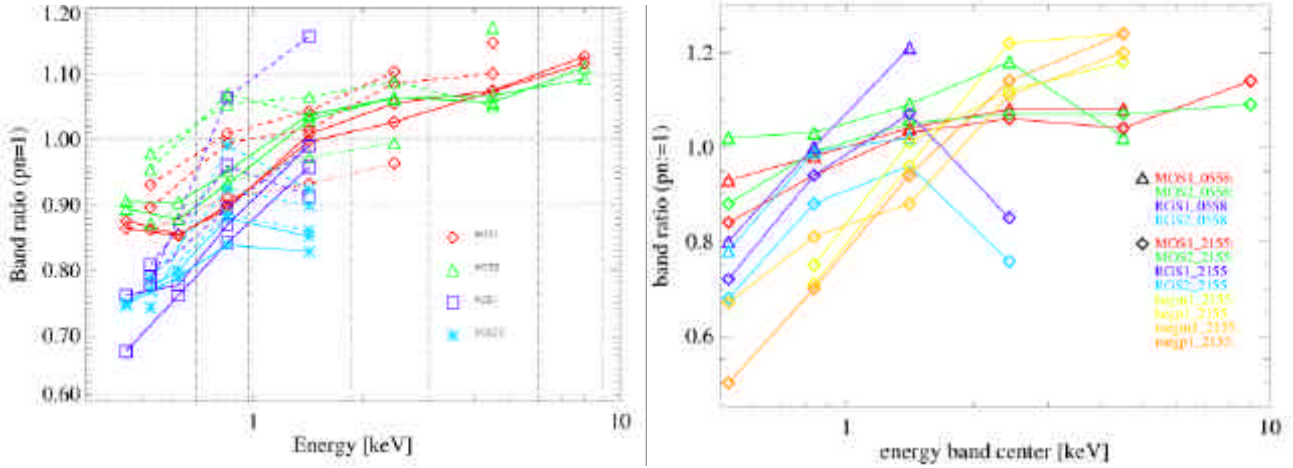

Figure 10: Left: normalisation factors for MOSs and RGSs with respect to pn on 4 different AGNs, Right: normalisation factors for MOSs and RGSs with respect to pn for PKS2155-304 and PKS0558-504 also including Chandra data

For spectral fitting over the whole energy range (0.2-12keV), EPIC MOSs are in relatively good agreement with pn, with a global normalisation very close to 1, within 2-3 %, slightly below for MOS1 and slightly above for MOS2. Although MOSs agree now significantly better, with SAS 6.0 and the latest PSF CCFs, both in slope and global normalisation, MOSs still show a trend for flatter (or harder) spectra, i.e. a trend for increasing MOS normalisation factors with energy. However the power law indices of EPICs are consistent within errors in the range 1-10 keV (?G < 0.07). The MOS/pn flux difference in the lowest energy bad (0.35-0.7 keV) is on average about 10-15 %. That might be partly attributed to inadequate modelling of the pn-redistribution at low energies, as in general RGSs tends to agree better with MOSs. But the discrepancy is rather variable and source dependent: for some sources the agreement is very good at low energies. For instance good agreement is observed between MOS1 and pn on E1821+643. On other sources the discrepancy increases to up to 15 %, in the 0.35-0.7 keV band. An extreme case in our analysis is the AGN MCG-6-30-15 where the pn excess in the energy band 0.35-0.7 keV is 20 %. The agreement tends to be better for faint sources, suggesting that pattern pile-up, or x-ray loading (in the pn offset maps) or a "lower threshold effect" for pn could be at play.

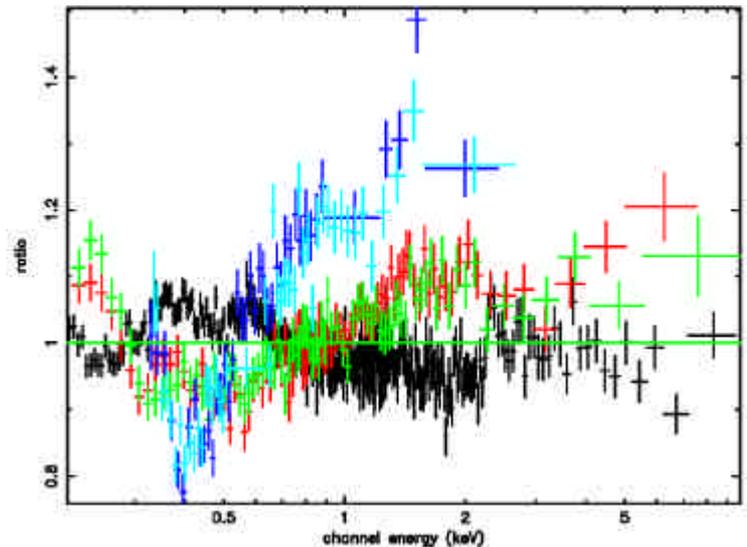

Figure 11: Ratio of data and model of simultaneous spectral fits to PKS2155-304. Black: pn, red, MOS1, green: MOS2, dark blue: RGS1, light blue: RGS2. Also here the new PSF correction improved the overall normalization from ± 12 % to ± 5 %. The larger residuals below 600 eV are due to uncertainties arising from the complicated shape of the redistribution function and the low energy effective area in combination with a gain problem of the pn-SW mode that is currently under investigation.

MOS2 shows a systematic higher normalisation factor of about 5 % relative to MOS1 below 2 keV, possibly pointing to a real effective area effect.

RGSs display in general a 20 % lower flux than EPICs in its energy range 0.35-1.84 keV. But as the discrepancy is higher in the lowest energy band (0.35-0.7 keV), RGSs tends to return systematically harder spectral slopes. However for some sources the agreement with MOSs is rather good above 0.5 keV, for instance in the MCG-6-30-15 case.

RGS2 tends to return lower fluxes than RGS1 above 1 keV, by about 10 %.

Fig. 10 shows also the ratios of the HETG (High Energy Transmission Grating) of Chandra for PKS2155-304 overlaid. The ratios are even more extreme than RGSs at low energy indicating also harder spectra, in closer agreement to RGSs than EPICs.

In Fig. 11 we show a typical example of the fit residuals for the different XMM instruments. The data have been fitted with an absorbed powerlaw where all parameters have been linked to the pn model allowing however an overall normalization.

All previously mentioned systematics can be seen:

- MOS fluxes lower by about 10-15 % below 0.7 keV with respect to pn.
- Lower fluxes by up to 20-30 % for RGSs below 0.7 keV with respect to pn.
- Harder spectral slopes returned by RGS

Note that with the good statistics of long exposures, systematic negative residuals can be seen for pn at the Silicon (1.8 keV) and Gold (2.2 keV) edges.

### 5.2 SNR

The comparative analysis of RGS and EPIC data of the SNR 1ES0102-7219 implied mode-dependent inconsistencies in the EPIC-pn energy scales (see Section 2.5). For this extended source, RGS data were selected from the whole RGS aperture with backgrounds synthesised from long-exposure blank fields according to the time-variable background rate monitor. EPIC source data were also selected from the whole of the projection of the RGS detectors onto the EPIC focal planes. This SNR's spectrum is unusually simple: because of an apparent total absence of Fe lines, a spectrum can be successfully modelled with 37 lines from CVI, OVII, OVIII, NeIX, NeX, MgXI and MgXII of which the O and Ne lines are particularly strong. With the RGS line positions and widths fixed empirically according to the RGS spectrum, a set of best fit line fluxes is calculated modulated by constant factors for each instrument in the four relevant XMM energy bands.

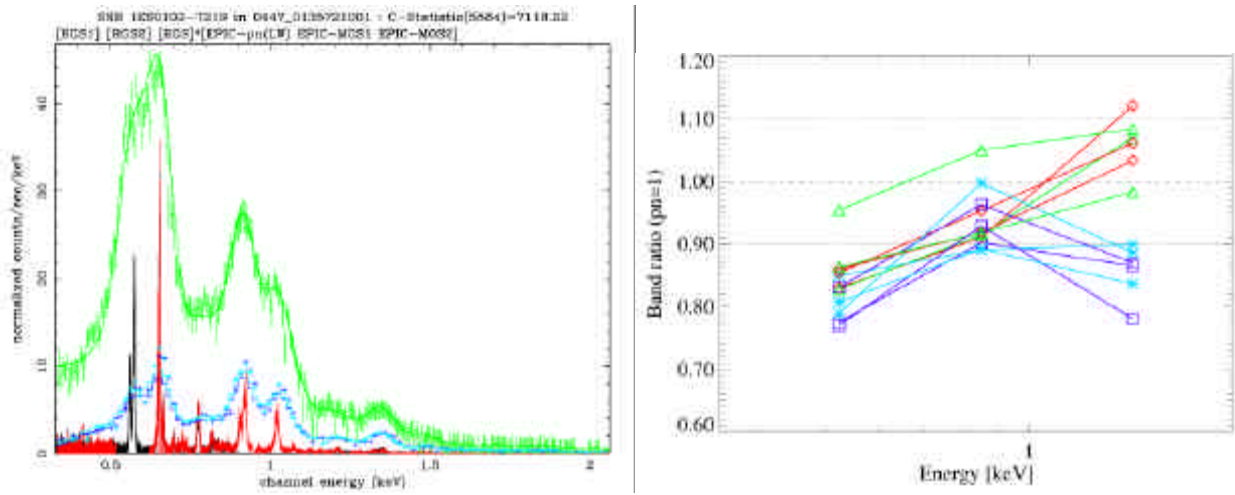

Figure 12: Left: 1ES0102-7219 observation from revolution 447 spectral fitting, with pn in large window. Black: RGS1, red: RGS2, green: pn, dark blue: MOS1, light blue: MOS2. Right: normalisation factors for MOSs and RGSs with respect to pn in different energy ranges: red: MOS1, green: MOS2, dark blue: RGS1, light blue RGS2.

The lines in EPIC-pn (FF) and EPIC-pn (LW) data both agree quite well with the RGS model lines as regards line position and width, although they clearly disagree slightly with each other: the RGS lines lie between LW and FF. As judged by the fit C-statistic, LW is marginally better. The SW data, on the other hand, have clearly been shifted to lower energies by about 20eV. This indicated a CTI under-correction that has since been improved (see section 2.5). As far as EPIC-MOS is concerned, the measured emission line profiles are more peaked than the corresponding redistribution model suggests. The energy-band ratios are shown in Fig. 12. RGS errors are typically 1 % and MOS errors 0.5 % respectively. The energy band ratios are rather consistent with the AGN analysis, showing the same systematics.

- RGS fluxes
    - 15 % lower than pn in the 0.35-1.84 keV band
    - 20 % lower in the 0.35-0.7 keV band.
- MOS fluxes ~10-15 % lower than pn in the 0.35-0.7 keV band.

### 5.3 Isolated Neutron Stars

Two isolated neutron stars (INS), RXJ0806.4-4123 and RXJ0720.4-3125 have been analysed. The absorbed black body model is a reasonable representation of the INS observations. Thus the model we use here for spectral fitting is *const\*wabs\*bb*, where *const* is the normalisation factor for each instrument. RXJ0806 is relatively faint and there are not enough counts in the energy band 0.7-0.972 keV. The difference of normalisation of the MOSs and RGSs with respect

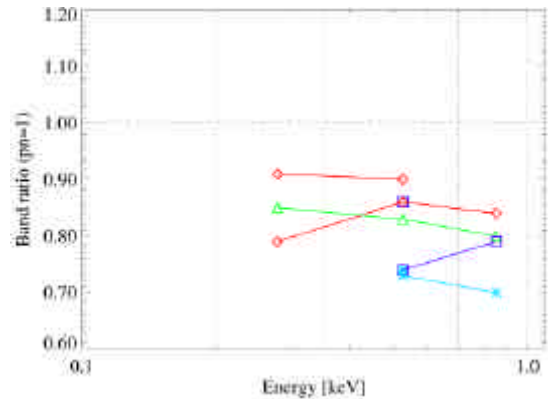

Figure 13: Normalization constants from different energy bands for the analysed isolated neutron stars from red: MOS1, green: MOS2, dark blue: RGS1, light blue: RGS2

to pn of about 20 % is illustrated in Fig. 13 where the normalization constants from different energy bands are shown. The most likely origin for the higher flux derived from pn is the improper redistribution at low energies (see section 2.4).

## 6. XMM-NEWTON VERSUS "THE REST"

Folding the RGS fluxed spectrum of the constant star ? Puppis through the detector response matrices of other instruments allows a direct comparison between them. Fig. 14 shows the spectra derived without any renormalization or other adjustment for XMM-Newton, ROSAT, ASCA and SAX. This RGS spectrum is mostly quite successful in reproducing the measurements with the exception of the EPIC-pn at low energies (see section 2.4)..

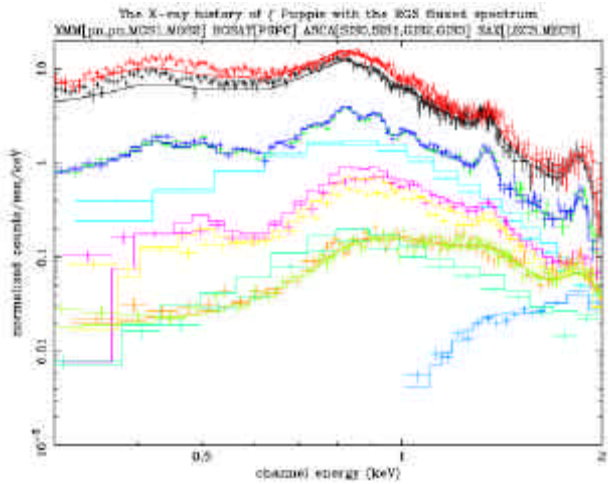

Figure 14: The RGS fluxed spectrum folded through various detector response matrices. From top to bottom pn (medium filter), pn (thick filter), MOS1&2, ROSAT PSPC, ASCA SIS0&1, ASCA GIS2&3, SAX LECS & MECS

## 7. POSSIBLE CROSS CALIBRATION IMPROVEMENTS

Various possibilities on how to improve the cross calibration differences are currently being discussed. We only list some of the calibration constituents, where there might be margin for adjustments: The effective area of the telescopes was determined by simulations and measurements before launch and could account for some effective area problems. An imperfectly modelled redistribution on both MOS and pn cameras could partly account for the low energy discrepancies. At the high end of the spectrum more margin is seen for adjusting the relative quantum efficiencies and also the overall absolute quantum efficiency for both cameras might need some adjustment. Some investigation has been put on to verification of the filter transmission that could partly explain the low energy differences. A major part of the disagreement of the effective area could also be accounted from the RGS, where the quantum efficiency could only be determined on ground with an error of around ± 15 %. All in all we seem to face a multi component problem that we shall tackle in the near future by reanalysis of ground measurements in combination with in-orbit calibration measurements in order to disentangle the various components.

## 9. ACKNOWLEDGMENTS

We thank the EPIC, RGS and OM PI teams for their efforts in calibration and their continuous support of the XMM-Newton calibration activities. The XMM-Newton project is an ESA Science Mission with instruments and contributions directly funded by ESA Member States and the USA (NASA). The German contribution of the XMM-Newton project is supported by the Bundesministerium für Bildung und Forschung Deutsches Zentrum für Luft- und Raumfahrt.